\documentclass[journal]{IEEEtran}
\usepackage{citesort}
\usepackage{amsmath}
\usepackage{epsfig}
\usepackage{subfigure}
\usepackage{enumerate}

\begin{document}

\title{On the Potential of Bluetooth Low Energy Technology for Vehicular Applications}

\author{\authorblockN{Jiun-Ren Lin$^1$, Timothy Talty$^2$, and Ozan K. Tonguz$^1$}\\
\authorblockA{$^1$Carnegie Mellon University, ECE Department, Pittsburgh, PA 15213-3890, USA\\
$^2$General Motors LLC, ECI Lab, Research and Development, Warren, MI 48092-2031, USA\\
Email: j.lin.us@ieee.org, timothy.talty@gm.com, tonguz@ece.cmu.edu}
}
\maketitle

\begin{abstract}
With the increasing number of sensors in modern vehicles, using an Intra-Vehicular Wireless Sensor Network (IVWSN) is a possible solution for the automotive industry to address the potential issues that arise from additional wiring harness. Such a solution could help car manufacturers develop vehicles that have better fuel economy and performance, in addition to supporting new applications. However, which wireless technology for IVWSNs should be used for maximizing the aforementioned benefits is still an open issue. In this paper, we propose to use a new wireless technology known as Bluetooth Low Energy (BLE) and highlight a new architecture for IVWSN. Based on a comprehensive study which encompasses an example application, it is shown that BLE is an excellent option that can be used in IVWSNs for certain applications mainly due to its good performance and low-power, low-complexity, and low-cost attributes. 
\end{abstract}

\IEEEpeerreviewmaketitle

{\small \textsl{\textcopyright2015 IEEE. Personal use of this material is permitted. Permission from IEEE must be
obtained for all other uses, in any current or future media, including
reprinting/republishing this material for advertising or promotional purposes, creating new
collective works, for resale or redistribution to servers or lists, or reuse of any copyrighted
component of this work in other works.}}

\section{Introduction}
Modern production vehicles are highly computerized, and the major functionalities of a vehicle are controlled by several Electrical Control Units (ECUs) inside the vehicle. ECUs need to gather information about the vehicle from the sensors in order to maintain all the required vehicular operations. Currently, most of the sensors inside vehicles are connected by physical wires, so each sensor sends out its data via the wires toward its destination ECU. However, because the complexity of vehicles is getting higher, and the number of applications and gadgets in vehicles keeps increasing, the large number of wires needed for the connection of sensors poses several significant challenges: the first one is the extra weight of the wires. If the extra weight can be eliminated, the weight of vehicles can be reduced and, thus, they can have better fuel economy and performance. Furthermore, the wired connection limits the possible sensor locations and hence the range of applications. The wires themselves are costly, and the cost for car manufacturers to install wires into vehicles can be high. When a vehicle gets older, some wires may deteriorate and cause severe problems, and to replace wires inside a vehicle would be either impossible or very expensive. In order to address these issues, wireless technology was recently proposed for the communications between sensors and ECUs. The wireless sensors and the ECUs form a new architecture, which is often referred to as an Intra-Vehicular Wireless Sensor Network (IVWSN)~\cite{GM-intra-vehicular}. 

Because of the potential benefits of IVWSNs, car manufacturers might gradually introduce wireless sensors into vehicles in the near future. The gradual scheme could start from several possible types of sensors: the ones which are not safety critical, the ones with hard-to-reach positions, or the ones which are the easiest to be replaced with wireless sensors. Furthermore, for car manufacturers, the additional cost of the wireless hardware is the major barrier for the deployment of IVWSNs. In order to massively deploy wireless sensors in vehicles, the unit price of a sensor with a wireless transceiver should not be much higher than an ordinary one. The cost of a wireless sensor highly depends on the chosen wireless technology and the complexity of the system. Consequently, a good starting point is to identify and evaluate a viable wireless technology to support the aforementioned types of sensors. These types of sensors/applications usually have the following requirements and properties:
\begin{itemize}
\item {\textbf{Requirements}}:
\begin{itemize}
\item{\textbf{\emph{Low Cost}}:} Lower complexity implies lower cost. Besides, if the system can adopt an existing wireless technology with minimum modifications, the cost can be further reduced. 
\item{\textbf{\emph{Low Power Consumption}}:} For most of the wireless sensors, their power is supplied by a battery. Therefore, the power consumption for the wireless communications has to be low enough to support a reasonable battery lifetime.
\item{\textbf{\emph{Short Delay}}:} For some of the applications, having a short delay (i.e., few milliseconds) is desirable since the system can be highly dynamic or requires prompt response. 
\item{\textbf{\emph{High Reliability}}:} The system has to provide guaranteed data transmissions.
\end{itemize} 
\item {\textbf{Properties}}:
\begin{itemize}
\item{\textbf{\emph{Low Data Throughput}}:} The sensor data are usually very short, i.e., only a few bytes.
\item{\textbf{\emph{Low Duty Cycle}}:} Most of the applications have a low duty cycle, e.g., less than 5\%. %
\item{\textbf{\emph{Various Priorities}}:} Depending on the applications, different packets are assigned with different priorities. For example, the packets from a safety-critical system generally have a higher priority than the packets from the air-conditioning system. %
\end{itemize} 
\end{itemize}
Moreover, while IVWSNs can be considered as a type of wireless sensor network, IVWSNs have a number of unique characteristics, and a specific protocol stack and system design would be required in order to achieve optimal performance. For instance, the sensors in IVWSNs are mostly fixed or can only move within a small area, while classical wireless sensor networks often have a dynamic topology~\cite{Tonguz06}. This implies that node mobility and routing configuration is less of a problem in IVWSNs. However, metal parts especially in the engine compartment act as obstacles and create a challenging and unique environment for wireless communications, especially compared to open space environments, as assumed in most of the classical wireless sensor networks. Due to the special physical environment, it is essential to evaluate the wireless technologies for IVWSNs in a bottom-up manner, starting from the Physical (PHY) layer. 

One of the wireless technologies that could be used for IVWSNs is ZigBee/IEEE 802.15.4 technology. Specifically, it has been shown before that ZigBee PHY layer is suitable for IVWSNs~\cite{ZigBee-based-WSN}. However, the investigation in~\cite{GM-intra-vehicular} has shown that the MAC protocol of ZigBee standard may not be suitable for some sensors/applications, and that could imply a customized protocol stack. In fact, since the sensors and applications in a vehicle are heterogeneous (i.e., with different requirements in terms of delay, throughput, duty cycle, and power consumption), it might be necessary to use more than a single wireless technology to fulfill all the requirements of different applications. The future IVWSN might therefore be a hybrid network with multiple wireless technologies coexisting for different groups of sensors. These factors necessitate further research into different options for wireless technologies.

Another possible wireless technology for IVWSN is ultra-wideband (UWB) communications. UWB was introduced in IEEE 802.15.4a-2007 and IEEE 802.15.4-2011 standards as one PHY option, and other variations of UWB have been studied extensively for the applications within a vehicle~\cite{bas2013ultra}. However, while UWB could provide a very large data throughput (e.g., several to hundreds of Mbps) and better resilience to multi-path fading, the cost of UWB technology is still higher than some existing low-power wireless technologies such as ZigBee. Furthermore, for automotive applications, the frequency spectrum of use has to follow the regulations worldwide, and this is one of the major reasons to use a low power wireless technology that operates in the 2.4 GHz ISM band which is available worldwide.

In this paper, we propose to use the Bluetooth Low Energy (BLE) technology~\cite{BLEspec} as an excellent choice for the IVWSN architecture\footnote{The Bluetooth\textsuperscript{\textregistered} word mark and logos are registered trademarks owned by Bluetooth SIG, Inc.}. The properties and the performance of BLE will be evaluated specifically for IVWSN applications. With our comprehensive evaluation and discussion, we show that BLE could provide a powerful hardware platform and PHY layer for IVWSN and enable car manufacturers to design and implement IVWSNs with low cost and high efficiency. %

The rest of the paper is organized as follows. Section II gives an overview of the BLE technology. Section III describes the IVWSN based on BLE and presents a detailed comparison between BLE and ZigBee. Section IV provides detailed information on the system design, configuration, and the methodology used for an example application: a BLE-based passive keyless entry system. Section V discusses the major issues related to the proposed system and applications. Finally, concluding remarks are given in Section VI.

\section{Overview of Bluetooth Low Energy}
\begin{figure}[tp]
\centering
\subfigure[Protocol Stack]{
\includegraphics[width=0.41\textwidth]{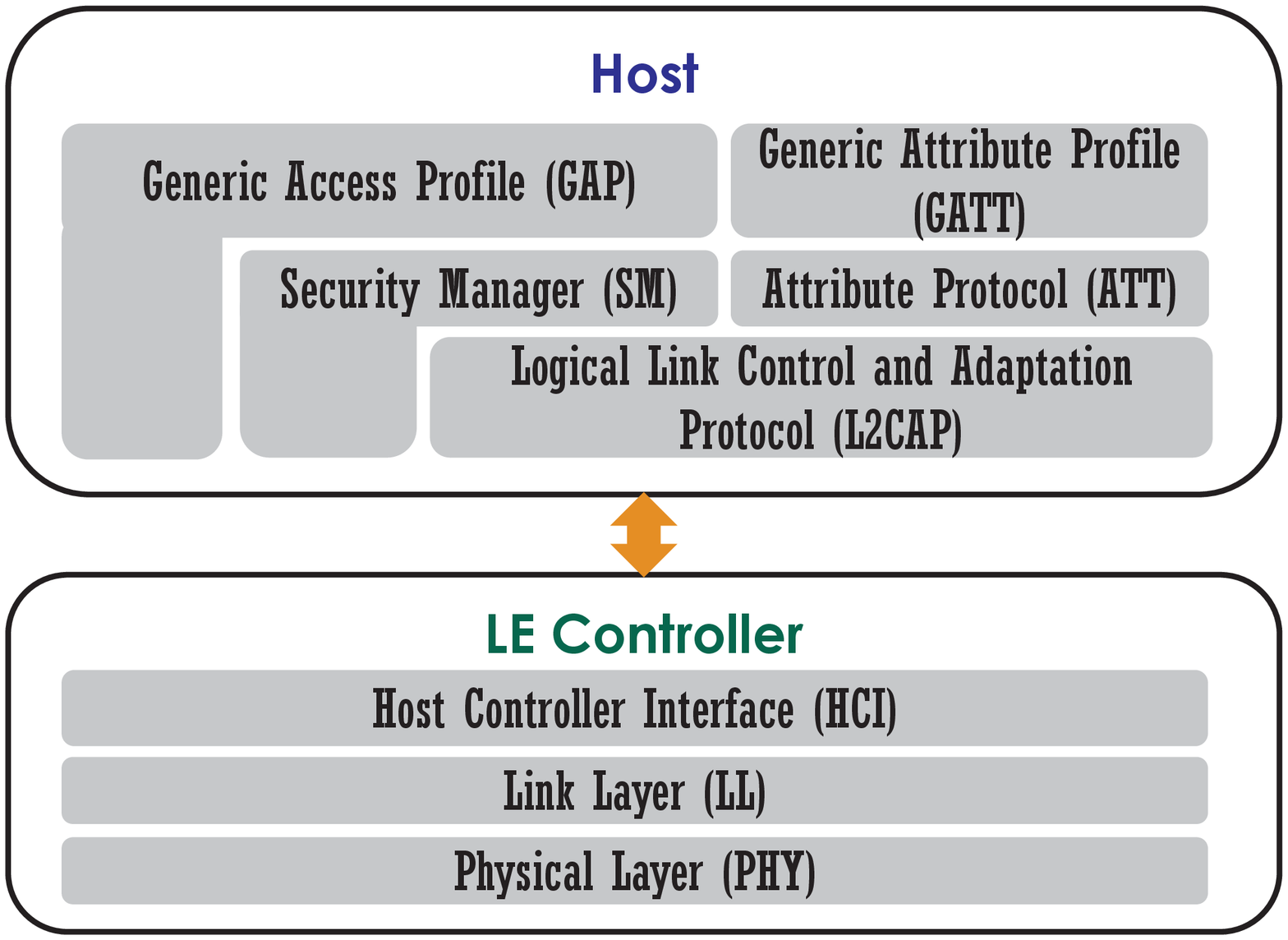}
\label{BLEstack}
}
\subfigure[Frame Format]{
\includegraphics[width=0.48\textwidth]{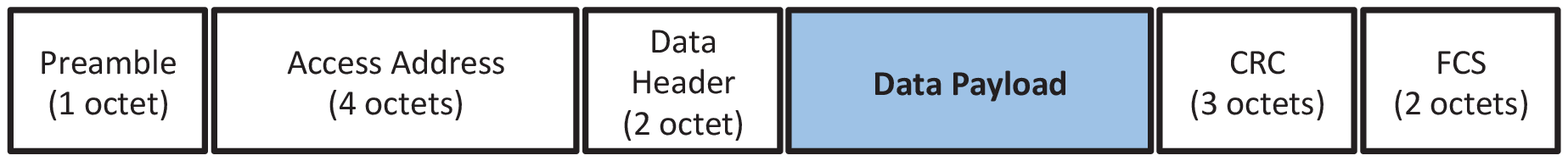}
\label{BLEframe}
}
\caption[]{The protocol stack and frame format of Bluetooth Low Energy}
\label{BLE_Std}
\end{figure}

Bluetooth Special Interest Group (Bluetooth SIG) announced the Bluetooth specification version 4.0 in June, 2010. It introduced the new Low Energy (LE) Core Configuration, which is also called Bluetooth Low Energy (BLE) in order to distinguish it from the traditional Basic Rate (BR) and Enhanced Data Rate (EDR) Core Configurations~\cite{BLEspec}. BLE is designed for applications which have low duty cycle and requires low power consumption and low cost. Fig.~\ref{BLEstack} shows the protocol stack of BLE. Note that the Bluetooth core system consists of a Host and one or more Controllers. 
A Bluetooth device could have both BR/EDR and LE controllers or only either one.%

BLE operates in the unlicensed 2.4 GHz ISM band, and it employs adaptive frequency hopping scheme to combat interference and fading. It uses 40 channels with center frequencies 2402 to 2480 MHz, and each channel is separated by 2 MHz. Among the 40 channels, three are advertising channels, and the remaining 37 channels are data channels. BLE uses binary Gaussian Frequency Shift Keying (GFSK) as the modulation scheme, and the symbol rate and bit rate are both 1 Mbps. The transmitting power of a BLE device is between -20 dBm to 10 dBm.

BLE has two different logical communication groups: one is piconet, and the other one is broadcast group. In a piconet, there is one master device and multiple slave devices\footnote{The maximum number of slaves in a piconet is not defined in the Bluetooth standard, but it is limited by the capabilities of the master device.}. All communications within a piconet is between the master and slave devices. There is no direct communications between the slave devices in a piconet. In other words, a piconet has a star topology. Before joining a piconet, a slave device can try to join a piconet by broadcasting advertisements on the advertising channels. The master device scans the advertising channels and decides if it wants to establish a connection with the advertising slave device. If the master device allows the advertising slave device to join the piconet, it will initiate the connection to the slave device. After the connection is established, the slave device is synchronized to the timing and frequencies of the physical channel specified by the master device. 
Note that in a piconet, each slave device uses a different physical channel (i.e., a different frequency hopping sequence) to communicate with the master device. 

On the other hand, a broadcast group consists of one advertiser and multiple scanners within the communication range of the advertiser. An advertiser broadcasts advertisements, and scanners scan the three advertising channels and receive the advertisements. There is no continuous connection between the advertiser and the scanners. In other words, while the master and slave devices are doing one-to-one connection-oriented communications in a piconet, the advertiser and scanners are doing one-to-many connectionless communications in a broadcast group. 

In a piconet, after the connection is established, there are periodic connection events between the master and each slave device. In a connection event, the master transmits packets to a slave and the slave can respond with a packet depending on the context. Therefore, the master controls the access to the channel in a piconet. Each connection event corresponds to a PHY hop channel. Consecutive connection events correspond to different PHY hop channels. The period of the connection events is defined by the upper layers.

In a BLE Host, the Generic Access Profile (GAP) layer (see Fig.~\ref{BLEstack}) controls the device's communication modes and procedures. Depending on the purpose of an application, the GAP layer operates in one of the following four roles: broadcaster (advertiser), observer (scanner), peripheral (slave), and central (master). In addition, a BLE device that operates in the peripheral or central role can also operate in the broadcaster or observer role. The application layer can control the operation role of the device by calling GAP API functions.

For packets in a connection event, each link layer packet uses a 24-bit cyclic redundancy error check (CRC) to cover the payload. If the CRC verification fails at the receiver, the packet will not be acknowledged and the sender will retransmit the packet. On the other hand, there are no acknowledgments or CRC field for the advertisement packets (broadcast packets). Each advertisement is transmitted several times to increase the probability that the scanner can successfully receive at least one of the copies. The length of a regular BLE packet is between 10 bytes and 47 bytes (as shown in Fig.~\ref{BLEframe}); the length of a BLE advertisement packet is between 8 bytes and 39 bytes.

The latest Bluetooth specification to date is version 4.1 which was announced in December, 2013~\cite{BLEspec41}. The major enhancement of the LE portion in Bluetooth version 4.1 is the additional link layer topology support. Bluetooth specification version 4.0 assumes that an LE slave device is only able to join one piconet at a time, but in Bluetooth version 4.1, an LE slave device can also act as a master or slave device of another piconet. Therefore, a scatternet topology is allowed in the new specification.

\section{IVWSNs based on Bluetooth Low Energy}

According to the existing literature, the intra-vehicular wireless channels have several properties~\cite{Moghimi09}:
\begin{itemize}
\item The 90\% coherence bandwidth at 2.4GHz is around a few MHz, which is at least as large as some indoor channels.
\item The coherence time of the intra-car channels ranges from 2.5 seconds to a few hundred seconds depending on different driving scenarios.
\item Huge path losses (e.g., $>$ 80 dB) can be observed when the transmitter and the receiver are in different compartments.
\end{itemize}
Along with the aforementioned requirements of sensors/applications inside vehicles, the candidate wireless technologies have to be low-power, low-cost, and occupy less than few MHz of bandwidth. As mentioned previously, BLE is designed for applications which have low duty cycle and requires low power consumption and low cost, and the channel bandwidth of BLE is 2 MHz, which is narrower than the coherence bandwidth inside the vehicle. These imply that BLE could also be suitable for IVWSNs as well.  

\begin{figure}[htp]
\centering
\subfigure[Packet goodput]{
\includegraphics[width=0.48\textwidth]{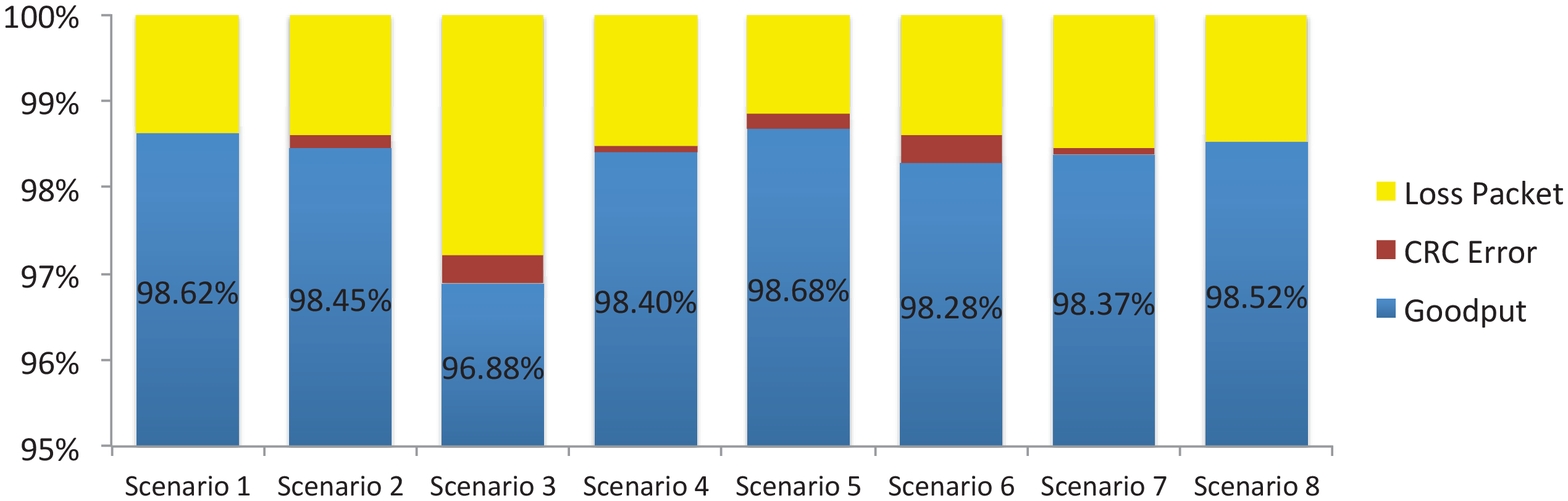}
\label{BLE_no_int}
}
\subfigure[Intra-vehicular scenarios]{
\includegraphics[width=0.45\textwidth]{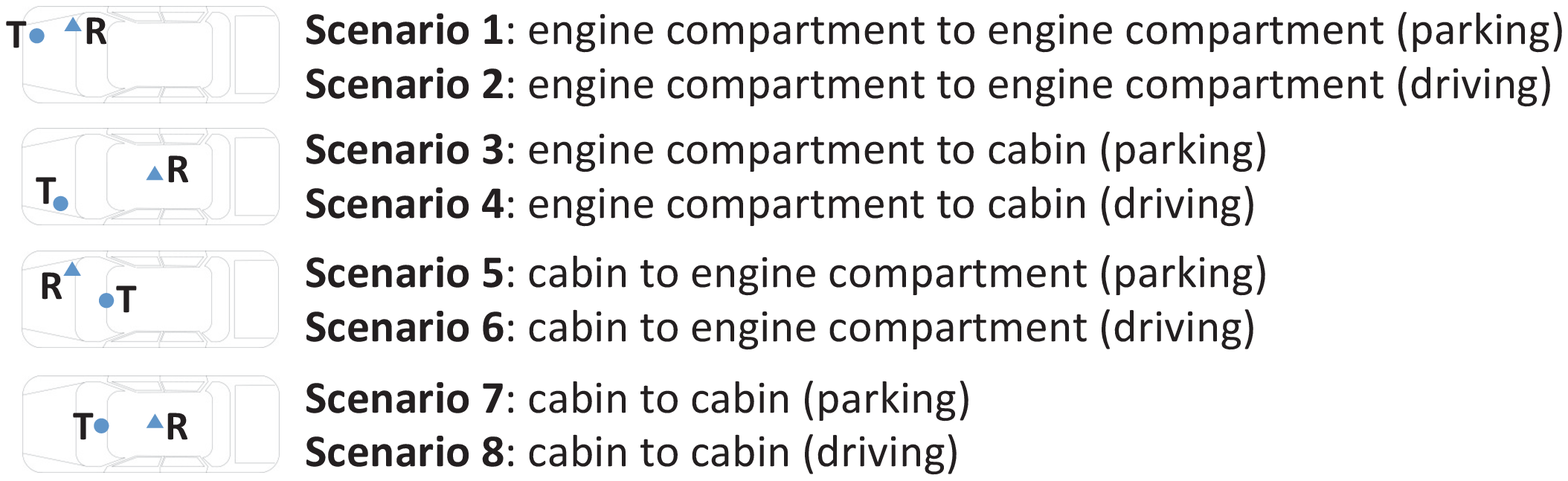}
\label{BLE_no_int_scenarios}
}
\caption[]{The packet goodput of BLE in eight different intra-vehicular scenarios}
\label{P_goodput}
\end{figure}

Since BLE was not originally designed for vehicular applications, we conducted a series of experiments in order to evaluate the actual performance of the PHY layer of BLE in an intra-vehicular environment. As part of the results reported in~\cite{Lin_Globecom2013}, it was shown that BLE can provide reasonably well packet goodput in the eight different intra-vehicular scenarios (see Fig.~\ref{BLE_no_int}). 
Fig.~\ref{BLE_no_int_scenarios} illustrates the positions of the BLE transmitter (denoted by \textbf{T}, the transmission power was 0 dBm) and receiver (denoted by \textbf{R}) in each scenario.

%

\subsection{Performance Parameters}
In addition to packet goodput, other important considerations for the wireless technology for IVWSNs are the throughput and delay performance. Since an IVWSN is mainly designed for sensor data communications, a large data throughput might not be required. However, if the technology provides more PHY layer throughput, the network will have a larger capacity to accommodate more sensors and data. This is important as car manufacturers are adding more and more features and sensors to modern vehicles. The data communications of BLE is performed in the predefined 37 data channels. The system can support multiple concurrent data communications if each master-slave pair applies an orthogonal hopping sequence. 
Theoretically, the maximum PHY layer throughput of the entire system could be up to 37 Mbps if all of the hopping sequences and traffic are carefully arranged.
Note that the actual data throughput would depend on the payload size of the sensor packets and the MAC scheduler design.
Compared to data throughput, the delay performance plays a more important role for many automotive applications. Delay is normally measured from the moment that a sensor sends out a data packet to the time the destination ECU receives the packet. For IVWSNs, some sensor data have to arrive at the destination ECUs within few milliseconds to maintain normal operation of the vehicle. The overall delay consists of three parts: transmission delay, queueing delay, and propagation delay. The transmission delay directly depends on the link data rate and packet size. Since sensor packets are usually fairly short, the major factor that affects the delay performance is the queueing delay. For instance, if the packet size is 20 bytes (i.e., with 8 bytes payload), the transmission delay is 0.16 ms since the data rate of BLE is 1 Mbps. The propagation delay is about a few nanoseconds depending on the dimension of a vehicle and hence could be ignored in most cases. In BLE, slave devices can only send a packet to the master device during the connection events after receiving a packet from the master. The queueing delay is the delay incurred while waiting for the connection event in order to send the sensor packet. Therefore, the connection event has to be carefully scheduled according to the sensor reading time in order to minimize the queueing delay.

\subsection{Comparison between Bluetooth Low Energy and ZigBee}

Several existing works on IVWSNs focused on ZigBee wireless technology~\cite{GM-intra-vehicular}~\cite{ZigBee-based-WSN}~\cite{tsai2007ICC}~\cite{Potential_intra-vehicular}. ZigBee is designed for RF applications which require low power consumption, low complexity, and low data rate~\cite{ZigBee2007}. The PHY and MAC layers of ZigBee are based on IEEE 802.15.4-2003 standard. Similar to ZigBee, Bluetooth, another Personal Area Network (PAN) technology, also operates in the 2.4 GHz unlicensed ISM band. According to the conclusions in~\cite{ITRLwirelessComp}, Bluetooth Basic Rate (BR) and ZigBee are both suitable for low data rate applications with limited battery power. However, Bluetooth BR still consumes more power and has higher complexity than ZigBee does. The main motivation for using BLE is therefore to provide a better solution for low-power and low-cost applications.

\begin{table*}[ht] \caption{The comparison chart of Bluetooth Low Energy and ZigBee} %
\centering %
\begin{tabular}{ccc} %
\hline\hline %
& Bluetooth Low Energy & ZigBee\\ [0.5ex] %
\hline %
IEEE Standard                &  None                                 & 802.15.4-2003                  \\
Frequency Band               &  2.4 GHz                              & 868/915 MHz, 2.4 GHz      \\
Max Data Rate                &  1 Mbps                               & 250 kbps                  \\
Nominal range                &  up to 50 m                           & 10 - 100 m                \\
Nominal TX Power             &  0 dBm                                & -25 - 0 dBm               \\
Number of RF Channels        &  79                                   & 25 (16 in 2.4 GHz)        \\
Channel Bandwidth            &  2 MHz                                & 0.3 / 0.6; 2 MHz          \\
Modulation                   &  GFSK                                 & O-QPSK                    \\
Spreading                    &  FHSS                                 & DSSS                      \\
Basic Cell                   &  Piconet                              & Star                      \\
Extension of the basic cell  &  None                                 & Cluster Tree, Mesh        \\
Max number of cell nodes     &  $>$65000                               & $>$65000                    \\
Data Protection              &  16-bit CRC                           & 16-bit CRC                \\
Connectivity                 &  supported by Bluetooth V4.0 devices  & dedicated devices         \\
Interference Avoidance       &  Adaptive Frequency Hopping Scheme    & Dynamic Channel Selection \\
Current Consumption (TX, 0 dBm output power)     &  TI CC2540: 21 mA                      & TI CC2430: 27 mA             \\
Current Consumption (RX)     &  TI CC2540: 15.8 mA                      & TI CC2430: 27 mA           \\
MAC Design                   &  Mostly TDMA                          & Flexible                  \\
Lowest Current Unit Cost     &  TI CC2540F128RHAR: \$2.59                                & TI CC2430F128RTCR: \$6.06                     \\ 
[1ex]
\hline %
\end{tabular} 
\label{BLEZigComp} %
\end{table*}

Table~\ref{BLEZigComp} is a comparison chart of BLE and ZigBee in terms of several important characteristics. Observe that they have many similarities: both of them operate in the 2.4 GHz ISM band, and the bandwidth of each channel is the same (i.e., 2 MHz).  However, since they use different modulation and spreading schemes, their maximum data rates are different: BLE can achieve up to 1 Mbps data rate, which is higher than ZigBee's 250 kbps. 
Another important advantage of BLE is the lower hardware cost. Both BLE and ZigBee are designed to be low-cost technologies, but the unit price of a BLE chip is currently less than a ZigBee chip. A possible reason might be that there are more phones and laptops supporting BLE as a part of the Bluetooth 4.0 standard, so it has a larger market than ZigBee does. It also implies that there will be more and more consumer devices which will support BLE in the near future, and it can enable new features on vehicles with lower cost.

Regarding the energy consumption, the current consumptions of a BLE and a ZigBee compliant chip are comparable. For example, the current consumption of Texas Instruments CC2540 BLE compliant chip is 15.8 mA and 21 mA for receiving (RX) and transmitting (TX), respectively~\cite{CC2540datasheet}. On the other hand, the current consumption of Texas Instruments CC2430 ZigBee compliant chip is 27 mA and 27 mA for RX and TX, respectively. However, the maximum data rate of ZigBee is 250 kbps, while BLE's is 1 Mbps. Even though the packet overhead is not considered yet, the normalized energy consumption
 of BLE would be smaller than ZigBee. Furthermore, because of the difference in their data rate, the transmission delays with BLE are smaller, and the delay performance can be very important for certain delay-sensitive vehicular applications.

Regarding reliability and robustness, BLE employs adaptive frequency hopping scheme to combat coexistence and fading problems, while ZigBee employs dynamic frequency selection. Under interference, BLE can dynamically update the frequency hopping sequence to exclude the channels with interference during active communications. ZigBee, on the other hand, selects a clearer channel before the communications starts, and then it sticks to the selected channel. Although ZigBee can choose to change channels periodically, it is still not as dynamic as BLE. As a result, BLE could be more sustainable over transient interference. We reported that when no interference exists in a car, the performance of BLE and ZigBee for IVWSNs is comparable. However, if strong WiFi interference is introduced, BLE can provide better performance than ZigBee~\cite{Lin_Globecom2013}.

Compared to BLE, ZigBee provides greater flexibility in terms of network topology and MAC design. For instance, the basic topology of a ZigBee network is star, but it also supports cluster trees or mesh. On the other hand, BLE only supports piconets (and scatternet if Bluetooth version 4.1 is used) in connection mode,
which follows a star topology. This, however, is not a problem for communications between ECUs and sensors\footnote{It is worth pointing out that many researchers are currently looking into replacing \emph{only} the wired links between an ECU and the sensors it is connected to with wireless links. 
Hence, with high probability, the network between the sensors and an ECU will have a star topology.
}. One can consider the ECU as the master device in a piconet, and the sensors as the slave devices. The ECU (i.e., the master device) coordinates the communications of sensors in the piconet. Regarding the MAC design, since ZigBee applies direct sequence spread spectrum, although the standard MAC protocols of ZigBee employ CSMA and TDMA, ZigBee can also use a large number of customized MAC protocols based on CSMA, TDMA, FDMA, CDMA, or a combination thereof. However, BLE can only use time-division (or reservation-based) MAC protocols due to the nature of frequency hopping spread spectrum. Therefore, for an IVWSN based on BLE, it is necessary to carefully design a scheduler for each piconet in order to accommodate the requirements of each sensor/application in the piconet. %

\section{Example Implementation}
\subsection{Experimental Platform}
\begin{figure}[tp]
\centering
\subfigure[Texas Instruments CC2540 Mini Development Kit]{
\includegraphics[width=0.35\textwidth]{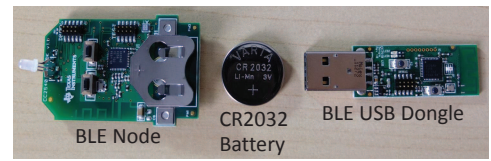}
\label{CC2540kit}
}
\subfigure[System diagram]{
\includegraphics[width=0.26\textwidth]{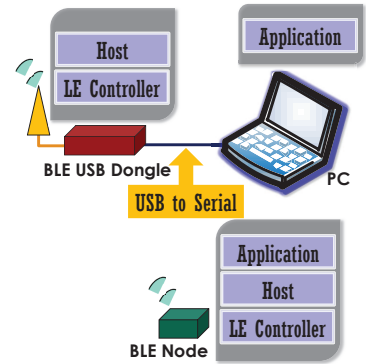}
\label{BLE_sys_d}
}
\caption[]{The Bluetooth Low Energy experimental platform}
\label{BLEplatform}
\end{figure}

The experimental platform used in this paper is based on Texas Instruments CC2540 Mini Development Kit~\cite{CC2540kit}. Texas Instruments CC2540 is a single-chip BLE solution which is capable of executing the BLE protocol stack and applications with a built-in 8051 microcontroller~\cite{CC2540datasheet}. The development kit includes a BLE node and a USB dongle, as shown in Fig.~\ref{CC2540kit}. The BLE node is powered by a CR2032 coin battery. The architecture of our experimental platform is depicted in Fig.~\ref{BLEplatform}. The USB dongle is connected to a PC with a USB to serial link. On the USB dongle, there are Host, LE Controller, and an adaptation layer which serves as the interface between the Host and the PC. The application layer and a serial port interface are implemented on the PC. On the BLE node, there are the application layer, Host, and LE Controller. Note that in the real automotive platform, the application layer will be implemented on an ECU (instead of a PC), and it can use a Universal Asynchronous Receiver/Transmitter (UART)
link to communicate with the CC2540 BLE chip.

\subsection{A Passive Keyless Entry System}
A Passive Keyless Entry System refers to a vehicle that can detect the key in its proximity and unlock itself (or unlock when the user pulls the door handle) when the key appears within a certain range from the vehicle. Several car manufacturers currently provide similar features on their production vehicles. However, in many of the current implementations (which usually use a low-frequency RF to detect the transducer on the key fob), the current consumption of the system on the vehicle could be high, e.g., around 700mA in some GM cars. To prevent draining of the battery, the system has to enter the sleep mode when it is idle, and it incurs undesirable long latency when the system gets reactivated.
To address the high current consumption and the long latency issues and provide a solution with lower cost, as a proof-of-concept, we have designed a Passive Keyless Entry System based on the proposed BLE IVWSN platform.

\begin{figure}[tp]
\centering
\subfigure[The components of Application Layer on the central device]{
\includegraphics[width=0.22\textwidth]{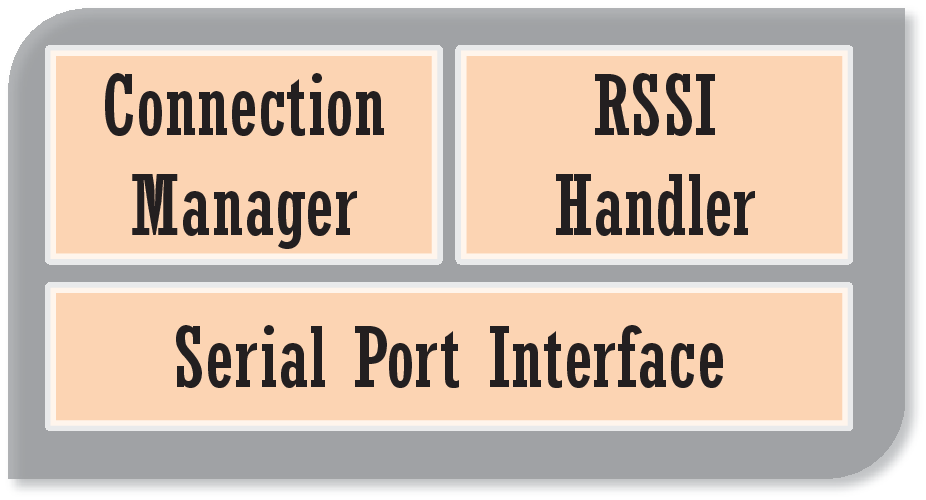}
\label{AppStack}
}
\subfigure[The state diagram of the connection manager component]{
\includegraphics[width=0.45\textwidth]{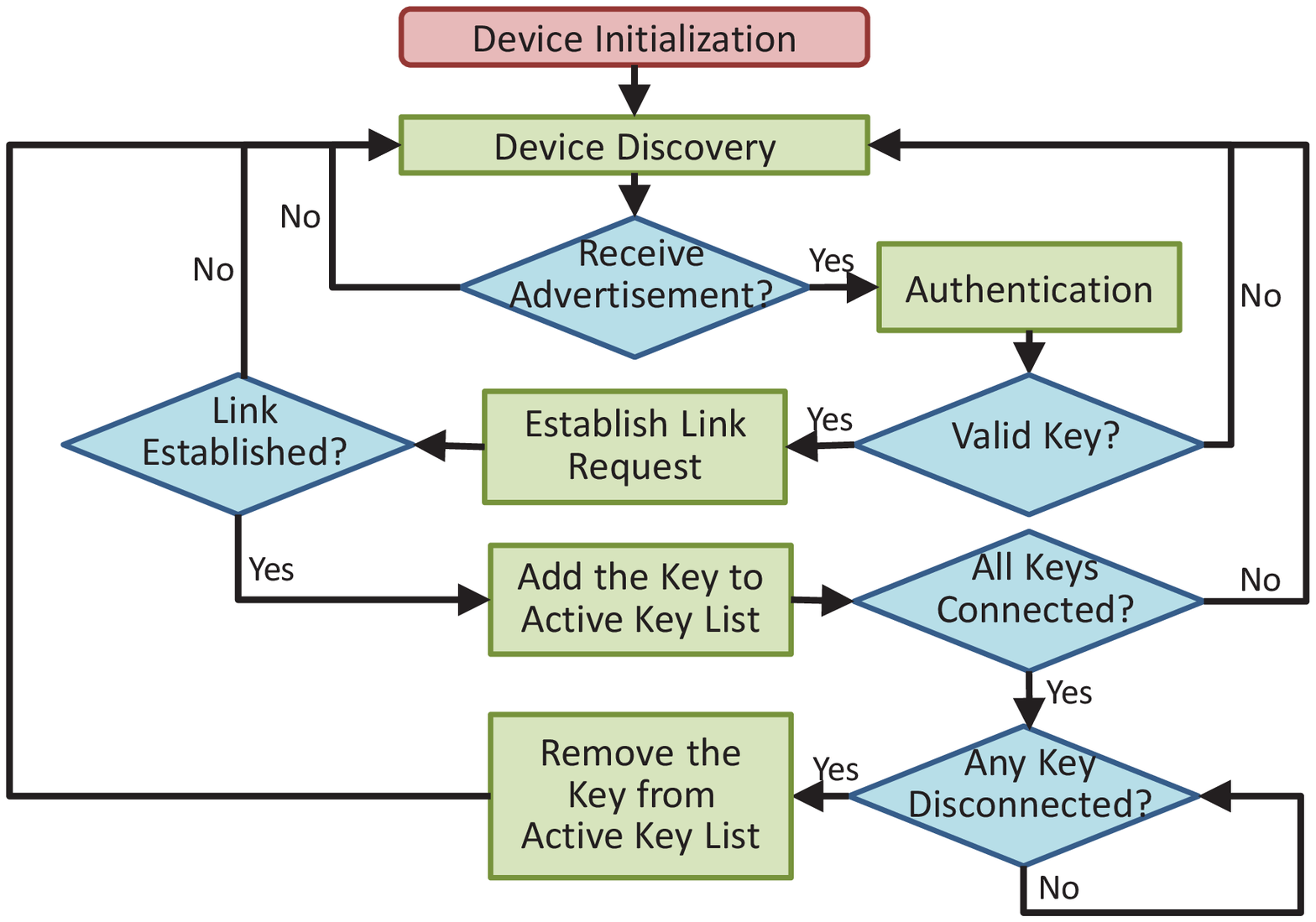}
\label{keyflow}
}
\caption[]{The application components of the Passive Keyless Entry System}
\label{app_comp}
\end{figure}

The test vehicle used in the experiment is a 2009 Cadillac STS. Two BLE nodes represent the BLE-enabled keys, and the USB dongle along with a PC is installed on the test vehicle to represent a lock control system on the vehicle. The keys are programmed as BLE peripheral devices. After powering on, the keys periodically send out advertisements with authentication information, and the keys will accept the connection if the connection request from the central device carries the correct pass code. On the vehicle, the USB dongle is programmed as a BLE central device, and its behavior is controlled by the application implemented on the PC. On the PC, there are three components in the application layer (as shown in Fig.~\ref{AppStack}). One of the components is the Connection Manager, which initiates and maintains the BLE connections to the keys; the other one is the RSSI Handler, which monitors the RSSI measurements of the packets from the keys and determines if the car should be unlocked.
The third component is a serial interface for communicating with the USB dongle. 

The flow chart of the Connection Manager is shown in Fig.~\ref{keyflow}. The Connection Manager maintains a valid key list and an active key list. The valid key list is pre-defined and should be pre-programmed by the car manufacturer. According to the valid key list, the Connection Manager scans for advertisements from those valid keys and handles the BLE connections to them. If any valid key is discovered, the Connection Manager will initiate the connection to the key and add the key to the active key list. 
After the BLE connection is established, the RSSI measurements are taken during each connection event. In each connection event, the central device sends a packet to the key, and then the key sends another packet back to the central device. The RSSI measurement of the latter packet is collected by the RSSI handler. According to the active key list, the RSSI handler collects the RSSI measurements from all of the active keys and determines if the system should unlock the doors when the user pulls the door handle.

\begin{figure}[tp]
\centering
\includegraphics[width=0.40\textwidth]{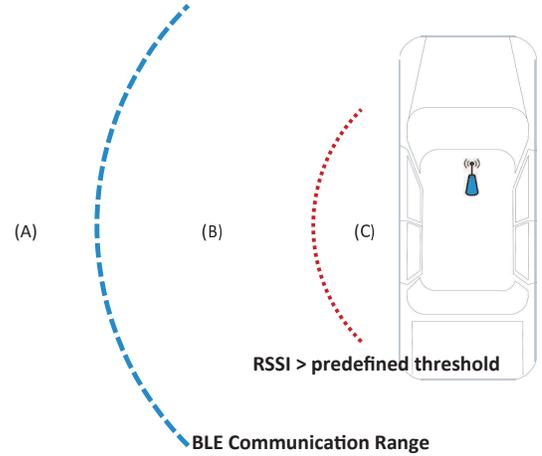}
\caption{The operation of the Passive Keyless Entry System based on BLE}
\label{keyless_op}
\end{figure}

As shown in Fig.~\ref{keyless_op}, the position of the key can be categorized into three regions based on the RSSI measurements. In region \textbf{(A)}, the key is out of the BLE communication range (e.g., around 25 m when the transmission power is 0 dBm); in region \textbf{(B)}, the key has an active connection to the central device; in region \textbf{(C)}, the key has an active connection and the RSSI of the packets from the key is larger than a predefined threshold (e.g., -55 dBm). Only when a valid key is in region \textbf{(C)}, the system will unlock the doors of the vehicle when the user pulls the door handle. Also, when there is no active key in the range for more than a certain period of time (e.g., 30 seconds), the system can choose to lock the car. 

The design was evaluated under the test case that the driver with the key walks toward and pulls the door handle, and then walks away from the vehicle 50 times. The system could correctly unlock the car every time the driver pulled the handle and the response time is negligible to the driver. This system can be easily integrated to the BLE IVWSN in future vehicles, thus providing a low-cost, low-latency, and highly-efficient solution for passive keyless entry system.

\section{Discussion}
In this paper, we demonstrated an IVWSN experimental platform based on BLE technology. In the future, there can be two possible ways to implement the BLE IVWSNs in production vehicles. In addition to replacing wired sensors with BLE sensors, the first way to set up the network is to install several standalone BLE central devices and attach them to the vehicle bus. The BLE central devices would serve as gateways between BLE sensors and the ordinary ECUs. The main advantage of this approach is the fact that changes to the existing architecture and other components would be minimal. However, the total cost of the related components in one vehicle would be higher. The other option is to add a BLE compliant chip or daughter board into multiple ECUs, so that the ECUs can directly communicate to BLE peripherals. However, this approach involves changes to the current ECUs and the initial cost and the effort needed to make such changes will be larger. Also, the positions of the ECUs and the placement of BLE antennas will be additional important design issues.

The other major issues are the MAC design and the channel capacity of the system when there are multiple BLE master devices existing in a single vehicle. As mentioned in the previous sections, BLE supports mainly time-division MAC protocols due to the nature of its PHY layer characteristics. Therefore, for the deployment of IVWSNs in a production vehicle, it is critical to calculate a schedule for all the sensors and ECUs to follow in order to achieve maximum performance and minimize interference. We are currently investigating the mechanism that determines the schedule.

It is important to note that BLE technology is fully equipped to protect the privacy and security of the communications. Encryption in BLE uses Advanced Encryption Standard in the Counter with Cipher Block Chaining - Message Authentication Code Mode (AES-CCM) cryptography, and multiple keys are generated by the host for data and device authentication. BLE also supports a privacy feature that can change the Bluetooth device address on a frequent basis to prevent a LE device being tracked by eavesdroppers. However, since one of the design goals of BLE is to keep the cost and complexity of a slave device to a minimum level, the association modes of BLE are not as sophisticated as those in BR/EDR. It has been reported that the key exchange during association could be compromised under certain circumstances~\cite{Ryan2013}. For IVWSNs, since all of the devices are preinstalled in the vehicle, one potential solution is to pre-define and store the cryptographic keys in the ECUs and sensors. Future work should look into different ways of further enhancing the security of the system.

An important concern about the BLE wireless sensors is the battery life. According to~\cite{TI_AN092}, the average current consumption of a BLE connection event is 10.655 mA, and the average duration is 2.348 ms. The current consumption during the sleep state is 0.9 $\mu$A. If the connection interval of the system is 2 seconds, the average current consumption can be calculated as:
\begin{align*}
	I_c &=\frac{(10.655\mbox{ mA} \times 2.348\mbox{ ms}+0.9\mbox{ $\mu$A} \times 1997.652 \mbox{ ms})}{2000\mbox{ ms}}\\
	&= 0.013\mbox{ mA}
\end{align*}
The typical capacity of a CR2032 coin battery is 230 mAh, so the estimated battery life is:
\begin{align*}
	T_b =& 230\mbox{ mAh}/I_c\\ 
	=& 230\mbox{ mAh} / 0.013\mbox{ mA} \\
	=& 17692 \mbox{ hours} \cong 737 \mbox{ days} \cong 2 \mbox{ years}
\end{align*}
Therefore, if the connection interval is 2 seconds and the BLE sensor node is connected all the time, its estimated battery life can be up to 2 years, which is quite respectable for most vehicular applications.

\section{Conclusion}
In this paper, we have proposed Bluetooth Low Energy technology as an excellent choice for Intra-Vehicular Wireless Sensor Networks (IVWSNs). 
An in-depth comparison between Bluetooth Low Energy and ZigBee in the context of IVWSNs is also provided and pros and cons of the two options are highlighted. Furthermore, we reported an example application to demonstrate a use case for the Bluetooth Low Energy experimental platform for intra-vehicular wireless communications. 
The main motivation for implementing a passive keyless entry system based on Bluetooth Low Energy is to reduce the response time, power consumption, and the cost of the existing system. 
Overall, our results show that Bluetooth Low Energy is a promising and viable wireless technology for IVWSNs and certain automotive applications that require low power and low cost solutions.

\ifCLASSOPTIONcaptionsoff
  \newpage
\fi

\bibliographystyle{IEEEtran}
\bibliography{BLE_jrlin}

\begin{IEEEbiographynophoto}{Jiun-Ren Lin}
received the B.S. degree in computer science and information engineering from National Chiao Tung University in 2004, the M.S. degree in computer science and information engineering from National Taiwan University in 2006, and the Ph.D. degree in electrical and computer engineering from Carnegie Mellon University in 2014. His research interests include computer networks, wireless networks and communications, personal communication systems, vehicular networks, algorithm design, performance evaluation, and machine learning. Dr. Lin is a member of IEEE and Eta Kappa Nu. He has served as the peer reviewer of several IEEE journals and conferences, including IEEE Transactions on Mobile Computing, IEEE Wireless Communications Letters, GLOBECOM, VNC, WCNC, SECON, ICC, VTC, and WONS.
\end{IEEEbiographynophoto}
\begin{IEEEbiographynophoto}{Timothy Talty}
received his B.S.E.E. from Tri-State University, Angola, Indiana, in 1987, and his M.S. and Ph.D. from the University of Toledo, Ohio, in 1990 and 1996. From 1982 to 1988 he was employed as a civilian with Naval Sea Systems Command, U.S. Navy, Arlington, Virginia. He joined Ford Motor Company in 1993, and worked on wireless channel modeling and concealed antenna systems development. He joined the EECS Department of the United States Military Academy, West Point, New York, as an assistant professor in 1997, where he conducted research on embedded antenna systems and high-speed Sigma-Delta converters. In 2001 he joined General Motors Corporation, Warren, Michigan, where he is currently a technical fellow working in the areas of wireless sensors and networks.
\end{IEEEbiographynophoto}
\begin{IEEEbiographynophoto}{Ozan K. Tonguz}
is a tenured full professor in the Electrical and Computer Engineering Department of Carnegie Mellon University (CMU), Pittsburgh, Pennsylvania. He currently leads substantial research efforts at CMU in the broad areas of telecommunications and networking. He has published about 300 papers in IEEE journals and conference proceedings in the areas of wireless networking, optical communications, and computer networks. He is the author (with G. Ferrari) of the book \textit{Ad Hoc Wireless Networks: A Communication-Theoretic Perspective} (Wiley, 2006). He is the founder, President, and CEO of Virtual Traffic Lights, LLC, a CMU spinoff that was launched in December 2010, which specializes in providing solutions to several transportation problems, such as safety and traffic information systems, using vehicle-to-vehicle and vehicle-to-infrastructure communications paradigms. His current research interests include vehicular ad hoc networks, wireless ad hoc and sensor networks, self-organizing networks, smart grid, bioinformatics, and security. He currently serves or has served as a consultant or expert for several companies, major law firms, and government agencies in the United States, Europe, and Asia.
\end{IEEEbiographynophoto}

\end{document}